\documentstyle[aps,preprint]{revtex}
\begin{document}
\title{First-Order Melting and Dynamics of Flux Lines in a Model for 
YBa$_2$Cu$_3$O$_{7-\delta}$}
\author{Seungoh Ryu\cite{ryuemail}}
\address{Department of Physics, Ohio State University, Columbus, OH 43210}
\author{D. Stroud}
\address{Department of Applied Physics, Stanford University, Stanford, CA94305
\\
and \\ Department of Physics, Ohio State University, Columbus, OH 43210\cite{stroud}
}
\date{\today}
\maketitle
\begin{abstract}
We have studied the statics and dynamics of flux lines in a model 
for YBa$_2$Cu$_3$O$_{7-\delta}\, $, 
using both Monte Carlo simulations and Langevin dynamics. The lines are 
assumed to be flexible but unbroken in both the solid and liquid states. 
For a clean system, both approaches yield the same melting curve, which is 
found to be weakly first order with a heat of fusion of about $0.02 k_BT_m$
per vortex pancake at a field of $50 {\rm kG}.$ The time averaged magnetic
field distribution experienced 
by a fixed spin is found to undergo a qualitative change
at freezing, in agreement with NMR and $\mu {\rm SR}$ experiments. The 
calculations yields, not only the field distribution in both phases, but
also an estimate of the measurement time needed to distinguish these 
distributions: we estimate this time as $\geq 
0.5 \mu{\rm sec}.$
The magnetization relaxation time in a clean sample slows dramatically as the
temperature approaches the mean-field upper critical field line $H_{c2}(T)$ 
from below.
Melting in the clean system is accompanied by a proliferation of
free disclinations and a simultaneous disappearance of hexatic order. Just 
below melting, the defects show a clear magnetic-field-dependent 
3D-2D crossover from long disclination lines parallel to the c-axis at low
fields, to 2D ``pancake'' disclinations at higher fields.  
Strong point pins produce an energy varying logarithmically with time.
This $\ln t$ dependence results from 
slow annealing out of disclinations in disordered samples.   
Even without pins,  
the model gives subdiffusive motion of individual pancakes in the dense
liquid phase, with mean-square displacement proportional to
$t^{1/2}$  rather than to $t$ as in ordinary  diffusion.  
The calculated melting curve, and many dynamical  features, agree well
with experiment. 
\end{abstract}
\newpage
\section{Introduction}
There have been numerous debates about the superconducting-to-normal 
transition in a magnetic field.
Abrikosov's classic mean field theory predicts a rigid vortex lattice
persisting to the mean-field transition line, $H_{c2}(T)$. More recent 
work suggests that the mean-field Abrikosov transition is altered by
fluctuations in both two and three dimensions (2D and 3D).  In the 2D case,
numerical evidence suggests
a first order melting transition from a vortex solid to
a vortex liquid\cite{teitel,hu,sasik95}, followed at higher temperatures by a
gradual crossover to a normal state.    
In 3D, the first suggestion of a first-order fluctuation-induced
transition is due to Br\'{e}zin {\it et al}\cite{brezin85}.
Their argument is based on Landau-level expansion of 
the Ginzburg-Landau functional 
for a conventional superconductor.

Much recent work on 3D flux-lattice melting transition 
has been stimulated by the behavior of high-T$_c$ 
materials\cite{blatter95,cubitt93,sllee93}.  
Several experiments have convincingly demonstrated
that the transition there is first-order in a sufficiently clean system at
sufficiently low fields\cite{safar92,zeldov95}.  On the theoretical side,
numerical studies of melting have been based on model pairwise
interactions\cite{xing90,ryu92,enomoto92,magro93,reefman93,probert95}, the
frustrated XY  model\cite{li91,hetzel92,shih84}, and an expansion of the free
energy in  lowest Landau levels (LLL)\cite{hu,sasik95}, among other approaches.
Some of these calculations\cite{sasik95,hetzel92} also indicate 
that the melting transition in the pure 3D system is first 
order, as suggested by experiment.
 
However, numerous issues remain unresolved.  
One such issue is the effects of disorder 
on the first-order transition.  
Disorder is widely expected to modify the
first order transition to either a vortex-glass\cite{fisher89} or Bose-glass
transition\cite{nelson92}, depending on the type of disorder. 
Modeling such disorder is difficult because of problems associated with
slow relaxation.  Much of the modeling has therefore been carried out
within the frustrated XY model\cite{reger91}, a model
which introduces an artificial
pinning by a fictitious lattice.  Other unresolved issues center on
the {\em dynamics} of the solid and liquid vortex system, for which only a 
very limited number of calculations have been carried out.  Dynamical
calculations are obviously necessary to understand many measurable properties
of the high-T$_c$ materials, such as the IV characteristics, voltage noise
spectra, NMR, and $\mu SR$.  A common approach  
is to model the high-T$_c$ material
as a network of Josephson junctions (a natural dynamical generalization of the
frustrated $XY$ model)\cite{lee93}. 
But as in the static case, this model also suffers from the problem of
fictitious pinning, though it is reasonably tractable numerically.
  
A few dynamical calculations have been 
carried out using a time-dependent Ginzburg-Landau (TDGL) model within
a vortex representation.  For example, Enomoto and coworkers 
have studied various aspects of flux lattice melting\cite{enomoto92} 
using this approach. They focused on the 
effects of pin density on the irreversibility line, using as a melting 
criterion the onset of flux line diffusion.  
They also considered the transport properties of pinned systems, 
but studied only configurations consisting of either a single flux line 
or a two dimensional lattice with point disorder.
Reefman and Brom applied a similar technique to a model for a single layer of
BiSr$_2$Ca$_2$Cu$_2$O$_8\, $, neglecting the Josephson coupling between layers and
focusing on the NMR properties\cite{reefman93}.  
Most recently, Probert and Rae\cite{probert95} 
have performed a Langevin simulation for a 
model YBa$_2$Cu$_3$O$_{7-\delta}\, $
system, both with and without pins.   Their results shed 
some light on the dynamical distinctions between the irreversibility line
(found in disordered systems) and the thermodynamic melting line 
(characteristic of a clean system).  Their calculations, however, assume
rigid vortex lines, thereby leaving out some of the most characteristic
three-dimensional (3D) behavior associated with flexible lines.
All these results suggest that this approach may be a useful and quite
realistic way to treat the dynamics of flux lines in high-T$_c$ materials.. 

In this paper, we present a numerical study of
both the statics and dynamics of a ``layered London model'' 
for a three dimensional flux lattice in  YBa$_2$Cu$_3$O$_{7-\delta}\, $, 
using a combination of 
Monte Carlo (MC) simulation and Langevin dynamics (LD) within a vortex 
representation. 
Our work extends earlier studies in a number of ways.  For example, we 
determine the melting line not only by the motion of individual
vortex lines, but also by changes in equilibrium quantities such as
the vortex structure factor and a hexatic order parameter.  We also determine
the conditions under which the vortex lines maintain their integrity, even
in the liquid state.  Perhaps of greatest interest, we find a novel
3D-2D crossover in the structure of the vortex solid just below melting,
at which the characteristic topological defects change from long disclinations
parallel to the $c$-axis to short disclination ``pancakes.''  This crossover
may possibly be connected with some recent experimental work, as discussed
further below.

The remainder of this paper is organized as follows.
The next section describes the model and discusses our choice of parameters
suitable for  YBa$_2$Cu$_3$O$_{7-\delta}\, $.  The following section presents our 
numerical results as obtained by both Monte Carlo and Langevin simulations. 
A brief discussion follows in the final section.

\section{Model}

\subsection{Model Classical Action}

In our model, interest is confined to fluctuations in the {\em phase} 
degrees of freedom of the superconducting order parameter $\psi$.
The amplitude 
$|\psi|$ is assumed not to fluctuate, but instead takes the value dictated by
minimizing the Ginzburg-Landau free energy at the given temperature
$T$ and magnetic induction $B$.  This resulting $|\psi|$ is related to 
an effective in-plane penetration depth\cite{lambda} by
\begin{equation}
\label{eqlambdadef}
\lambda_{ab}^2(T,B) \equiv \frac{m^*c^2}{4\pi|\psi|^2e^{*2}}
= \frac{\lambda_{ab}^2(0)}{\left( 1 - (T/T_c(0))^4 \right) 
\left( 1 - (B/B_{c2}(T))^2 \right)},
\end{equation}
where $T_c(0)$ is the mean-field transition temperature at zero magnetic field,
$\lambda_{ab}(0)$ is the in-plane penetration depth at zero temperature, and
$B_{c2}(T)$ is the mean-field upper critical field line.
$|\psi|$ is normalized so that $m^*$ is twice the electron mass $m_e$; and
$e^* = 2e$.

The model consists of $N_z\,$ parallel superconducting layers 
a distance $d$ apart.  Each layer contains $N_v\,$ two-dimensional 
``vortex pancakes'' (i.\ e., 2D vortices) 
described by transverse position coordinates ${\bf r}_{i,k}$. 
The vortex density is assumed fixed at $n_B \equiv 1 / a_B^2 \equiv B / \phi_0,$
where $\phi_0 = hc/2e$ is the flux quantum.
Such pancakes in different layers 
interact via both magnetic and Josephson interactions
\cite{lawrence70,carton91,clem90,feigelman90}.
Here, we simply assume that these interactions combine to produce
flexible but {\em unbreakable} vortex lines - that is,
each pancake is always associated with two specific pancakes in the adjacent 
layers.  The justification and possible limitations of this assumption are 
discussed below.
The interlayer coupling strength is characterized by a single variable
$\gamma = \xi_{ab}(0)/\xi_c(0)$,  where $\xi_{ab}(0)$ and $\xi_c(0)$ 
are the zero-temperature superconducting coherence lengths
in the ab-plane and c direction.
$\gamma$ has associated with it a length scale
$r_g \equiv \gamma d$. 
The layered structure becomes important 
for lengths shorter than $r_g$\cite{doniach89}.

Following \cite{ryu92}, we write down the Hamiltonian for the
system as
\begin{equation}
\label{eqaction}
{\cal H} = \sum_{i\neq j}\sum_k U \left(\frac{|{\bf r}_{i,k} - {\bf r}_{j,k} |}
{\lambda_{ab}(T,B)}\right)
+ \sum_{i}\sum_k V \left( {|{\bf r}_{i,k} - {\bf r}_{j,k} | \over  2 r_g} \right).
\end{equation}
Here the in-plane repulsive interaction takes the form
\begin{equation}
\label{eqko} 
U(x) = {d \phi_0^2 \over 8\pi^2 \lambda_{ab}(T,B)^2}K_0^* (x), 
\end{equation}
while the interlayer interaction is taken as
\begin{eqnarray}
\label{eqjos}
V(x) &=& c_J (x -1) \quad (x > 1); \nonumber \\
&=&  c_J (x^2-1) \quad (x \le 1),
\end{eqnarray}
with
\begin{equation} 
c_J = {d \phi_0^2 \over 8 \pi^3 \lambda_{ab}(T,B)^2}\left[ 1 + \ln \frac{\lambda_{ab}(0)}
{d}\right ].
\end{equation}
In order to reduce finite size effects, we
employ periodic boundary conditions in all directions.  Because of these, the
effective in-plane interaction becomes $K_0^*(x)$,
which represents the summation of the modified Bessel function 
$ K_0(x) $ over image vortices\cite{image}.

\subsection{Langevin Dynamics}

To probe real time dynamics, one can also run LD simulations on the same 
model, assuming that the vortices are subject to an overdamped dynamics. 
Then the equation of motion for a vortex pancake can be written
\begin{equation}
\label{eqlang}
\eta \dot{\bf r}_{i,k} (t) = {\bf f}_{i,k}^T(t) + {\bf f}_{i,k}^U(t) + 
 {\bf f}^L + {\bf f}^P ({\bf r}_{i,k} ).
\end{equation}
The first term on the right-hand side is the Brownian force
due to thermal noise.   The noise is assumed to be Gaussian-distributed
white noise with correlation functions
\begin{equation}
\langle f_{i,k}^T(t)\rangle = 0,
\end{equation}
\begin{equation}
\langle({\bf f}^T_{i,k}(t) \cdot \hat{n}_\alpha)\, ({\bf f}^T_{j,k'} (t') \cdot
\hat{n}_\beta) \rangle = 2 k_B T {\eta \over d} 
\delta_{ij}\delta_{\alpha,\beta} \delta_{kk'} \delta( t-t'),
\end{equation}
where $\hat{n}_{\alpha}$ is a unit vector in the $\alpha$ direction, 
$\alpha = x,y.$
The second term on the right-hand side of eq.\ (\ref{eqlang}) is 
the force due to 
the other pancakes; it is obtained as a negative
spatial gradient of the vortex-vortex interaction term, 
as written down in eqs.\ (\ref{eqko})-(\ref{eqjos}). 
The third term (not studied numerically in the present paper) is the 
Lorentz force due to an applied current.  

The last term in (\ref{eqlang}) 
describes the force due to the random pinning potential. 
The pins are modeled as uniformly cylindrical regions of 
radius $r_p$ [taken to be $2\xi_{ab}(0)$ throughout this work]. 
The pinning energy of the vortex pancake is expected to be 
determined by the  fraction of the core area within the pinning well.
We can achieve this dependence by assuming a pinning energy per pancake
$U_p(T,B) = \alpha_p d \phi_0^2 /[  16 \pi^2  \lambda_{ab}^2(T,B)]$.
Thus, the strength of a single pin is controlled by the 
dimensionless parameter $\alpha_p$.  The effectiveness of the pins is, 
of course also
influenced by their areal density $n_p$, or the ``equivalent field''
$B_p \equiv \phi_0n_p.$ 

For simplicity, the force due to the $l^{th}$ point pin is assumed to be
directed radially inward towards
its center (at ${\bf R}_l$). For $r_p > \xi_{ab}(T)$ it is given by
\begin{eqnarray}
f^P_l({\bf r}_{i,k}) &=& 
-{U_p(T,B) \over 2 \xi_{ab}(T)} \quad  \text{if}\quad  r_p - \xi_{ab}(T) < |{\bf r}_{i,k}
- {\bf R}_l | < r_p + \xi_{ab}(T), \nonumber \\ &=& 0 \quad \text{otherwise.} 
\end{eqnarray}
For $r_p < \xi_{ab}(T),$ it takes the form
\begin{eqnarray}
f^P_l({\bf r}_{i,k}) &=&  
- {U_p(T,B) \over r_p +  \xi_{ab}(T)} \Big({r_p \over \xi_{ab}(T)}\Big)^2  \quad
\text{if} \quad  0 < |{\bf r}_{i,k} - {\bf R}_l | < r_p + \xi_{ab}(T) \nonumber \\ &=& 0
\quad \text{otherwise.}
\end{eqnarray}
This choice includes in the simplest manner 
the fact that the vortex core area grows with  increasing $T$ while the
disorder is temperature-independent. 

\subsection{Numerical Approach and Choice of Parameters}
To obtain the thermodynamics 
via MC, we use the standard Metropolis 
algorithm with variable step sizes, as discussed in \cite{ryu92}.
Typically, we equilibrate over $2\times 10^4$ MC steps and evaluate 
the thermodynamic averages over an additional $2\times 10^4 -10^5$ 
steps. 

For both MC and Langevin calculations, we use look-up tables for 
both the potential and the forces, as well as a scheme for interpolating
between the points in the table. 
The time iteration is carried out using a second-order Runge-Kutta algorithm
in time steps of $\triangle \cdot t_0$ where $t_0 = {\phi_0 \over B}{d \eta
\over (32)^2 \epsilon_d}$ and  $\epsilon_d = {d \phi_0^2 \over 8 \pi^2
\lambda_{ab}(0)^2}$.  
The choice of $\triangle$ is dictated by the dominant forces
in our  model [eq.(\ref{eqlang})], 
which in this paper are the vortex-vortex interactions; in
general, $\triangle \le {\cal O}(10)$.  
For optimum convergence, we have allowed 
$\triangle$ to depend somewhat on $T, B$, and $J$,
since different components of the force may dominate at different values of
these parameters.

Finally, we briefly discuss our choice of parameters. 
For $\lambda_{ab}(0)$, we use 1000 $\AA$ for  YBa$_2$Cu$_3$O$_{7-\delta}\, $ 
single crystal. This
choice is close to the experimentally determined value, and also
places the simulated melting curve in close
agreement with the  experimental data of \cite{safar92}.
For the remaining parameters, we use the values
appropriate for  YBa$_2$Cu$_3$O$_{7-\delta}\, $:
$\kappa = 87.5$, $d = 11.1 \AA,$ 
$\gamma = 5$, $T_c(0) = 93 K$, and $dH_{c2}(T)/dT = -1.8 \times 10^4 {\rm 
Oe/K}$,
where $\kappa \equiv \lambda_{ab}(0)/\xi_{ab}(0)$ is the Ginzburg-Landau
parameter.
 
\subsection{Calculated Quantities}

Before discussing our numerical results, we first define a few important 
physical quantities.  $\delta R(t)$ is the transverse 
root-mean-square (rms) displacement of a pancake vortex from its
average position, i.\ e.
\begin{equation}
\label{eqdr}
\delta R(t) \equiv \frac{1}{a_B}
\left\{\frac{1}{N_{tot}}\sum_{i,k}
\left<({\bf r}_{i,k}-<{\bf r}_{i,k}>_t)^2\right>_t^{1/2}\right\}.
\end{equation}
Here $<\ldots >_t$ denotes an average over time t, and 
$N_{tot} = N_z N_v$ is the total number of pancakes.
Now in a finite system, as in our simulation, 
the collection of vortex lines tends to drift as a whole even in the 
solid phase.  This behavior, seen in both solid and liquid phases, is strictly
a finite-size effect and has no relation to any measurable quantities.
We therefore subtract out the drift by using 
${\bf r}_{i,k}^*(t) = {\bf r}_{i,k}(t) - {\bf R}_{cm}(t)$ instead of 
${\bf r}_{i,k} $ in Eq.~(\ref{eqdr}), ${\bf R}_{cm}(t)$ being the instantaneous
center-of-mass coordinate of the entire lattice. To monitor 
lateral fluctuations, we also 
calculate the ``wandering length'' $l_T$\cite{hellerqvist94} defined by
\begin{equation}
l_T^2 \equiv \frac{1}{N_{tot}a_B^2}
\sum_{i,k} \Big< |{\bf r}_{i,k} - {\bf r}_{i,k+1}|^2 \Big>_t.
\end{equation}
In addition, we compute the density-density correlation function
\begin{equation}
C({\bf r},z) =  <\rho_v({\bf r},z)\rho_v({\bf 0},0) >_{t\rightarrow \infty},
\end{equation}
and its partial Fourier transform 
\begin{equation}
S({\bf q},z) = \int d^2 r C({\bf r},z) \exp (i {\bf q} \cdot {\bf r}),
\end{equation}
where $\rho_v({\bf r},z)$ is the local vortex number density in each plane $z$  
and ${\bf q} = (q_x, q_y).$

\section{Numerical Results}

\subsection{Location of Melting Curve; Heat of Fusion}

To locate the melting point for a given $B$ by MC simulation, 
we first make a quick sweep over a wide range of temperatures
($\sim 20 K)$, using $2\times 10^4$ MC steps for each T in steps of
$\Delta T \sim 0.5-1 K$.  We interpret a discontinuous jump in $\delta R(t)$  
as well as the vanishing of the intensity of the Bragg
peak $S({\bf q},0)$ at ${\bf q} = {\bf G}_2$ (where ${\bf G}_2$ is a
2D reciprocal lattice vector of the triangular lattice)
as signatures of melting. 
We then repeat more careful sweeps over a
narrower temperature region with up to $10^5$ MC steps.

If the temperature is cycled through $T_m$ using an interval
of $10^4-3 \times 10^4$ MC steps for each temperature in 
increments $\Delta T/T_m =
0.0024$,  we observe hysteresis in most monitored quantities.
The width of the loop is typically $\approx 0.018 T_m$.  Hysteresis
is most pronounced for $\delta R(t)$ and for the disclination density
(defined below), but
is also quite conspicuous for the hexatic order parameter(also defined below)
 in the same
temperature range.  From the size of the jump in the total internal energy seen
at melting, we estimate the latent heat per vortex pancake to be about 
$0.034 \pm 0.01 k_BT_m$.  (A possibly more precise estimate is given below.)
The melting transition is calculated to occur at 
$T_m(B) /T_{c2}(B) =0.87 \pm 0.02, 0.93 \pm
0.006$, and $0.93 \pm 0.02 $, for $B = 90$, $50$, and $10$ kG respectively,  
and lattices with eight layers, 
in reasonable accord with experiment\cite{safar92}.
For lattices with 32 layers and $\lambda_{ab} = 1400\AA$, we obtain
$T_m(B) /T_{c2}(B) =0.86 \pm 0.02, 0.92 \pm
0.02$, and $0.96 \pm 0.02$ at the same fields.  
[Here $T_{c2}(B)$ is the mean-field transition
temperature at field B.]   
Since our  assumed T-dependence of $\lambda_{ab}(T,B)$ is likely to become less
accurate as $T\rightarrow T_c(0),$  
we may expect increasing deviations from experiment
at lower fields, as indeed we find in our calculations. 
Note that, for the higher fields, $T_m/T_{c2}(B)$ depends little on 
the either the lattice aspect ratio or the value of $\lambda_{ab}$. 
This behavior is consistent with the dimensional crossover discussed
below. 

The transverse displacement $\delta R(t)$, as expected,  
shows characteristically different behavior in the solid and liquid phases.
In the liquid, $\delta R(t)$ increases with increasing $t$ (see below), 
while in the solid phase, it saturates after a short transient period.  
The inset to Fig.\ 1 shows the behavior of $\delta R(t)$ 
across the melting transition for $B = 50 {\rm kG},$ as calculated by MC.  
Also shown  are the Langevin results, which agree very well with MC predictions. 
This confirms that the two indeed give, as expected, 
 very similar predictions for 
thermodynamic
quantities. From the results in the
solid phase, we can read off the
Lindemann number $c_L \equiv \delta R(T_m, B) = 0.18$, at B$=50{\rm kG}.$
We have not, however, confirmed that $c_L$ is constant along the melting 
curve, as would be expected if Lindemann's law is really valid.

Another way to display the first-order melting process
is also shown in Fig.\ 1.   The MC energy at any given temperature fluctuates
with MC time.  The two curves  in the main part of Fig.\ 1 show 
the distributions of 
total internal energy within two different time windows at the melting 
temperature [(a) and (b) in Figs.\ 1 and 2].  On either
side of $T_m,$ 
this distribution tends to be independent of the initial time of the 
window.  Precisely at T$_m$, however, the system slowly oscillates between
a liquid and a solid phase with a correlation time of about $5\times 10^4$
 MC time steps (for this system size).  
The oscillation causes the two distributions to differ:
the two distributions shown are the extreme
cases that we have found in the length of time studied.  
As shown in the insets at the
bottom of Fig.\ 2, the
density correlation functions in the ``low-energy'' and ``high-energy''
windows do indeed show solid-like
and liquid-like characteristics.  We believe that the weak hexagonal
symmetry seemingly present  in the ``liquid'' phase actually
results from the presence of a few solid configurations in the
window, and possibly also from the finite size of the sample.

>From the average energy difference
between the two phases as read from Fig.\ 2, and also from the distance 
between the peaks in  Fig.\ 1, we estimate 
the heat of melting to be  
$\stackrel{<}{\scriptstyle \sim} 0.02 k_BT_m$ per vortex pancake.  
This estimate for the heat of fusion is in
agreement with LLL calculation of \cite{sasik95}, as well as with experimental
estimates at high fields.  It is significantly smaller, however, than 
Hetzel {\it et al}\cite{hetzel92}'s value of $0.3k_BT_m$ per
vortex pancake, obtained using a frustrated stacked triangular $XY$ model.  
The reasons for this
difference are a matter of speculation.  One possibility is that in both
our simulations and the LLL model, $|\psi|,$ 
which sets the overall energy scale, decreases with increasing temperature,
whereas in the XY simulation, the coupling strength $J$ is
temperature-independent.  More plausibly, our simulations are carried out
at high fields while the $XY$ simulations are more appropriate at low fields.
Yet a third (unlikely)
possibility is that our simulations allow for anisotropy in the
superconducting properties, whereas those of Hetzel {\it et al} do not.

We find that most of the
energy discontinuity at melting comes from changes
in the interaction energy between different pancakes in the same layer, which
has a rather clear jump at melting.  By contrast, the interlayer coupling
energy has large fluctuations in both the solid
and liquid phases but no clear jump. These fluctuations tend to mask
the step-like change of the in-plane
component.

\subsection{Topological Defects in Clean  YBa$_2$Cu$_3$O$_{7-\delta}\, $}

We have carried out a search for topological
defects above and below melting in our model for  YBa$_2$Cu$_3$O$_{7-\delta}\, $.  
Basic building blocks for various kinds of topological defects
are disclinations and dislocations. 
Both kinds of defects may be identified by carrying out a Delaunay
triangulation\cite{preparata85} for individual layers of 
each sampled configuration.
To understand this procedure, we show in Fig.\ 
\ref{figdefects} a snapshot of a typical melted configuration in a single layer
which shows both {\em bond lines}, as identified by the Delaunay procedure, 
and {\em topological defects}.  The vortex pancakes are located at the vertices 
of the triangles.  They are marked by black and grey dots 
if their number $n(i)$ of in-plane neighbors is seven (positive disclination)
or five (negative disclination) rather than the six expected for a 
perfect triangular lattice.  We see characteristic examples of an
isolated dislocation [i.e.\ a pair of disclinations, marked by (a)], 
an isolated disclination (b), and a bound pair of dislocations (c). 
In the present work, we arbitrarily call a 
pair of disclinations ``bound'' if and only if they reside on
neighboring pancakes as in (a) and (c).
 
For a more quantitative analysis of the defect configurations, we
assign ``charges''
$q(i,z)= n(i,z) - 6$ to the $iz^{th}$ pancake, where $n(i, z)$ is
the number of in-plane neighbors of the $iz^{th}$ pancake.
The {\em average} total defect density is defined by 
$n_d = {1 \over N_v N_z } \sum_{i,z} n_q(i,z)$ where 
$n_q(i,z) = 1 - \delta_{q(i,z),0}$.  (Note that this definition includes
defects of both signs.)
Since tightly bound neutral pairs of disclinations,
such as (a) and (c), do not influence local hexatic order (i.\ e.\ the degree
of local six-fold symmetry), it is also useful to 
define an {\em isolated} charge density by
$q^*(i,z) = q(i,z) + \sum_{j}^{\prime} q(j,z)$, where the prime indicates that
$j$ runs over the in-plane neighbors of pancake $(i,z)$. 
This definition eliminates bound-pair defects such as (a) and (c). 
The {\em average} isolated charge density is then defined by $n_d^* = {1
\over N_v N_z } \sum_{i,z} n_q^*(i,z)$, where $n_q^*(i,z) = (1 -
\delta_{q^*(i,z),0})$. 

Once the charges are identified on each plane, the topological
defect configuration associated with a given vortex arrangement
(lower left in Fig. \ref{figdefects}) 
may be represented as ``neutral plasma" of line charges of 
variable lengths $l_d$ and both signs (cf.\ Fig. \ref{figdefects},
lower right). To quantify the properties of this plasma, 
in our simulation, we monitor the distribution of defect lengths of either
sign ${\cal P}(l_d)$, defined as
\begin{equation}
\label{eqlzdef}
{\cal P}(l_d) = {1 \over N_z N_v} \Big< \sum_{i,z}\sum_q [1-n_q(i,z_0)][1 -
n_q(i,z_0+l_d+1)] \prod_{z=z_0+1}^{z_0+l} n_q(i,z) \Big>_t \/ .
\end{equation}
We also monitor the in-plane
hexatic order parameter $\Psi_6$, defined by 
\begin{equation}
\label{eqhex}
\Psi_6 = {1 \over   N_v N_z } \sum_{i,z} {1 \over n(i,z)} \sum_j
< \exp ( 6 i \theta_{ij}(z) ) > 
\end{equation}
where $\theta_{ij}(z)$ is angle made by the bond
between vortices  $i$ and $j$ and the  $x-$axis. The results are shown in 
Fig.~\ref{fighex}.
Evidently, melting is 
accompanied by a proliferation of isolated dislocation and 
disclination lines, as well as by a dramatic drop in the in-plane 
hexatic order parameter.  Near
$T_m$, but still below it, transient line defects are observed to 
appear and disappear, while at $T_m$ there is an 
abrupt decrease in hexatic order accompanied by an almost discontinuous 
jump in $n_d^*$ (cf. Fig.~\ref{fighex}).  
$|\Psi_6|$ appears to vanish no more than 1 \% 
above $T_m$.  While
we can not rule out a ``hexatic line phase'' (with hexatic
but no long-range crystalline order) within this temperature range, our
numerical results are consistent with a single melting transition
where both hexatic and crystalline order simultaneously disappear.
 
As the flux line density increases, the interlayer coupling becomes 
weaker compared to the in-plane interaction.   Hence, one might expect 
a dimensional crossover in the defects associated with the melting transition.
Indeed, we find such a crossover in the solid phase.
In Fig. \ref{figdefectcrossover}, 
we show the length distribution ${\cal P}(l_d)$ for 32 layers at several 
fields ($B=10,50,90$ kG) and temperatures slightly below $T_m$($T/T_m = 
0.997,0.982,0.988)$ selected with the   
criterion of $\delta R = 0.15$ for consistency.  
In each case, 
by monitoring $C({\bf r},z)$ and 
$S({\bf q}, z)$, we verified that the system is solid, 
but very near melting. 
As evidence that the lattice anisotropy depends on field, 
we note that for the same value of in-plane rms fluctuations $(\delta R =
0.15)$, $l_T$ is field dependent: at fields
of 10kG, 50kG, and 90kG, it is respectively 0.077, 0.100, 
0.114.  

The topological defects {\em have a dramatic cross-over} 
as a function of field.  This is obvious from the right hand column of
the Figure.  At the lowest field ($10 kG$),
there are line defects penetrating all the way through the sample, which occur
only for temperatures extremely close to melting($T/T_m > 0.99$).  
At higher fields, the defect line segments are
much shorter and occur at somewhat lower values of $T/T_m$.  
(The latter fluctuations
set in at a lower temperatures because they are short
and cost less energy to create.)
Also shown in gray are the defect length distributions for
three temperatures above the melting line, namely $T/T_m = 1.01, 1.04,
1.06$ for B$=10$, $50$, and $90 kG$ respectively.  
In this case there is no 3D-2D crossover: the defect length
distributions in these line liquid states are similar for all three fields,
following a simple exponential form. 
[However, at the still higher temperatures  
$T/T_m \stackrel{>}{\scriptstyle \sim} 1.05, 1.07,
1.13$ respectively (not shown), 
$l_T > 0.2$ and the lines begin to break up into
pancakes, as cutting and reconnections set in (see  
Section \ref{cuttingsection}).]

To account qualitatively for these results, we note that the 
defects shown in Fig.
\ref{figdefectcrossover} are line segments of isolated 
disclinations of various lengths.  The energy $E_c$
to create an isolated ``pancake'' (i. e. 2D) disclination 
defect, as is well known, is proportional to the logarithm of the 
system area.  It may be written approximately as
$E_c \approx J$ln$(L^2/a_B^2)$, where $L$ is the system edge, $a_B$ is
the lattice constant of the 2D lattice, and $J$ is some appropriate energy,
which is of the order of the intervortex interaction energy.  The energy to
create a line of $\ell$ such disclinations is therefore
(neglecting momentarily the interactions between the individual pancakes)
of the order of $E(\ell) \approx \ell E_c$. 
Since the total number of such defects in thermal equilibrium should be 
proportional to $\exp (-E(\ell)/k_BT)$, this argument will give the kind
of exponential distribution seen numerically in both the solid and liquid
phase.

To further refine this argument, we first note that $1/a_B^2$ is proportional
to the magnetic induction $B$.  Hence, for a given area and fixed $J$, 
the energy to create a 2D disclination increases as $ln B$, which implies that
the slope of the exponential dependence should become {\em steeper} as $B$
increases.  This increase is indeed observed in the solid phase, but it
seems to be much more abrupt than the gradual increase suggested by this
argument.  Presumably, the abruptness stems from interactions between
the disclination pancakes.  If this interaction energy increases exactly
as $\ell$, the exponential form would be unaltered.  Our numerical results
show some deviation from strictly exponential behavior at low fields, 
suggesting that the interaction energy is more complicated.  We speculate
that this interaction energy is stronger at low fields but is reduced at
larger fields, possibly by screening from other disclinations, leading to
the rather sudden transition to short defects seen at high fields.  But a
quantitative theory of this transition remains to be developed.

Why is there no such transition in the liquid state?  A possible but 
speculative explanation is that, in the liquid, 
there are many disclinations (of the order of 0.5 per plaquette
per layer of the vortex lattice).  Hence, they strongly screen one another,
and the effective sample area $L^2$ in the expression for the creation energy
should be replaced by the area of a plaquette,
which is of order $a_B^2$.  Then the energy to create a line of $\ell$ 
disclinations is independent of field (except possibly through the prefactor
$J$), implying a field-independent distribution, as observed
numerically.

Although the above arguments are certainly speculative, the principal numerical
result - namely, a rather sharp ``3D-2D'' crossover in the defect structure,
may have an experimental analog.  
Obara {\it et al}\cite{obara95} have recently reported a crossover
in multilayers of 
${\rm DyBa_2Cu_3O_7}$/${\rm (Y_{1-x}Pr_x)Ba_2Cu_3O_7}$, 
in which a 3D vortex lattice showed only 2D correlations above $\sim 10kG$.   
In the experimental sample, of course,
the vortex lattice is affected by point pins, possibly producing a sharper
crossover than we see here.  Nonetheless, these pins should
affect the vortex lattice quite differently at low and high fields, because
the defect lines  are clearly much less rigid at high fields\cite{ryu96}.  
Specifically, because
of this low rigidity [as shown in ${\cal P}(l_d)$], the point pins may 
cause the defect lines to break into short segments rather abruptly at a
well-defined magnetic field which may be speculatively 
identified with the transition observed by \cite{obara95}.

\subsection{Distribution of Magnetic Induction in Solid and Liquid.}

Both the static and dynamic magnetic field 
distribution are often probed 
experimentally\cite{sllee93,recchia95_1,pennington95,song95}, using techniques
such as $\mu SR$ and $NMR$. To calculate this distribution, we have 
evaluated the
instantaneous local magnetic field
$B({\bf r},t)$ at a given time $t$ (either by MC or LD),
using Clem's prescription\cite{clem91} for the field of a stack of
vortex pancakes.  In the present case, we have an 
additional complication due to the periodic
boundary conditions.   This difficulty is again
solved by including the effects of image pancakes both within the
plane and along the c-axis. 
 
To obtain the dynamic evolution of the field distribution, using LD, we
consider the {\em instantaneous distribution function}
\begin{equation}
\label{eqpbinst}
{\cal P}(B, t) = \frac{1}{V}\int dV \delta(B({\bf r}, t) - B),
\end{equation}
which will in general depend on time, as the system approaches equilibrium.
For a measurement which probes the local field averaged
over a time $t$, we obtain the field distribution
function ${\cal P}_t(B)$ from the following definition:
\begin{equation}
\label{eqpbt}
{\cal P}_t(B) = 
\frac{1}{V}\int dV \delta(\langle B({\bf r},t')\rangle_t -B),
\end{equation}
where $\langle \ldots \rangle_t$ denotes an average over a  time $t$ (either
over real time, for a LD simulation, or Monte Carlo time). From the limit
$t\rightarrow \infty$, we obtain the static field distribution function
appropriate for $\mu$SR,
(in a typical $\mu$SR, the muons sample the B-field
at random points in a sample averaged over a typical muon
lifetime of $\sim 10^{-6}$ sec.)
using either MC or LD. A similar technique has been
recently applied to obtain the static field distribution in BiSr$_2$Ca$_2$Cu$_2$O$_8\, $ using
MC\cite{schneider95}. By varying the duration $t$ over which the local field is
accumulated, we also find out how rapidly the field distribution approaches 
the static limit. 

Fig. \ref{figpbt} shows 
${\cal P}_t(B)$ as calculated using LD for $t/t_0$ ranging from $10^3$ to
$5\times 10^5$ and three different temperatures above and below melting.
In the long time limit, the solid phase exhibits 
the characteristically 
asymmetric profile arising from the static triangular lattice. 
In the liquid, the profile
becomes nearly symmetric (and very narrow) because vortices move around, 
producing the same time-averaged field everywhere. 
The distribution in the solid phase is qualitatively similar to 
that detected by $\mu$SR experiments in  
BiSr$_2$Ca$_2$Cu$_2$O$_8\, $ \cite{sllee93}.   Our symmetric result in the liquid
appears to disagree with the experiment.  
The discrepancy may result from the fact that our simulation sample, in
contrast to the experimental one, lacks a boundary.  The possible influence
of such boundaries on the field distribution observed by $\mu SR$ experiments
has been noted and discussed by Schneider {\it et al}\cite{schneider95}.
For short time scales ($t/t_0 < 10^4$), our
distribution retains significant asymmetry even in the liquid phase, suggesting
that the time-scale for field relaxation is of this order in the liquid.

Fig. \ref{figbsq} shows the change in rms width of the field distribution
across the melting point. In the long-time limit, the temperature dependence
of the width as obtained from both MC and LD agrees qualitatively with
NMR linewidth measurements across
melting\cite{recchia95_1}. This agreement suggests that the
time scale of the NMR measurement is longer than $10^4-10^5 t_0$.
Since the NMR experiments are carried out using a.c. fields of frequency 
$\sim$ MHz, this implies $10^4 t_0 \ll 1 \mu{\rm sec}$ or $t_0 \ll 10^{-10}
{\rm sec}.$  [This agrees with another estimate of $t_0$ given below.]
Note also that our interpretation of NMR linewidth neglects the possibility
that the a.c. NMR field actually exerts a force on the vortex lattice. 

Two more pieces of information may be extracted from these results.  First,
the inset in the Fig. \ref{figbsq} shows 
that a MC result taken over $4\times 10^4 $ MC steps gives an
rms width closely matching the LD result obtained with $t/t_0 = 10^5.$ 
This suggests that a single MC time step using our version of the
Metropolis algorithm with 
individual step size $\triangle x = a_B/32 $ is equivalent to 
$\approx 2.5 t_0$ of LD time. 
Also, from the LD results with variable $t/t_0$, we can  
infer that melting has a pronounced effect on the field distribution only when
the measurement is made on time scales longer than $\sim 10^5 t_0$.
In fact, the required time scale to distinguish solid from liquid
may be somewhat longer than even this
value in the thermodynamic limit.  
For the LD cell size used in these magnetic field calculations,
the individual pancakes may have $\delta R(t) \propto t^\alpha$ with 
$\alpha \sim 1/3$ rather than $\alpha \sim 1/4$ as expected for very long 
lines (see below).  Thus, for these thicker samples, 
we expect that the relevant
time scale for field relaxation should be closer 
to $(10^5)^{4/3} t_0 \sim 0.5 \mu$sec.

\subsection{Relevance of Vortex Line Cutting and Reconnection}
\label{cuttingsection}

Before turning to further dynamical results, 
we first consider the validity of
neglecting vortex line cutting and reconnection.  In an earlier numerical study
of BiSr$_2$Ca$_2$Cu$_2$O$_8\, $ using a similar model\cite{ryu92}, it was found that
including line-breaking effects had little effect on the calculated
melting properties.  In simulations of  YBa$_2$Cu$_3$O$_{7-\delta}\, $, which has far
stiffer lines,  it is reasonable to expect that the approximation is even
sounder.  Indeed, even in the liquid phase up to $T/T_{c2}(B) \sim 0.97,$ 
our numerical results show that the
transverse ``wandering length'' $l_T$,
measured in units of the intervortex separation, is no larger than $0.2$.

To check this approximation another way, define a 
``vortex collision length'' $\zeta_z$  by  
$l_T^2 (\zeta_z / d)^{\delta } \sim 1 $.  
Presumably $\delta = 1$ in the dilute (low-field) limit, where the 
transverse wandering of a line directed along the c-axis 
is a random walk with step size $l_T$.  Using 0.2 for $l_T,$ we obtain 
$\zeta_z/d \sim 25$,
which exceeds the thickness of our sample. 
In the dense regime, $\delta$ may be smaller than 1 because of the restrictive
effects of the repulsive interactions among flux lines, leading to an even 
larger $\zeta_z/d$.  

Cutting and reconnection should occur massively only
when the collision length becomes comparable to the interlayer spacing, 
leading to frequent ``collisions" of vortex lines.
By balancing the entropic gain from permutations of 
vortex connections against the accompanying cost in interlayer coupling
energy, we estimate that this condition should be met only when $l_T > 0.7$, 
which occurs only well above melting. 
To verify
this, we did two MC runs, in one of which we allowed cutting and reconnection 
according to a Boltzmann weight factor obtained from the change in
interlayer coupling energy that would be produced upon cutting and
reconnection.  The result shows that this cutting occurs at a negligible
rate until
$T/T_m > 1.05 $, at which point about 12\% of
recombination attempts are accepted. 
$l_T$ at this temperature was found to be about 0.16, a value which may
roughly be taken as a kind of ``Lindemann melting criterion'' for flux cutting
in the liquid state.  Thus, over much of the liquid
regime, we conclude that the 
thermodynamics of this model can be treated without
considering flux line breaking and reconnection.  This conclusion justifies our
treatment of flux line dynamics in the same approximation. 

\subsection{Langevin Dynamics of Vortex Line Liquid and Solid; Slow and
Fast Relaxation}

Having justified our non-breaking assumption, we return to the dynamics of 
vortex line liquids and solids. 
We begin by considering the motion of a {\em single} vortex line within
the overdamped dynamics of eq. (\ref{eqlang}).  In the limit of a very long 
line in which each pancake is 
subject only to thermal Langevin noise and a harmonic interlayer restoring
force, eq. (\ref{eqlang}) can be solved analytically.  The derivation is
similar to that of Ref.~\cite{blatter95} for more general cases.
The result for the  mean square displacement of a pancake from its 
initial position at time t is 
\begin{equation}
\label{eqdrdiff}
\langle |{\bf r}_{i,k}(t) - {\bf r}_{i,k}(0)|^2\rangle \approx
k_BT\left(\frac{t}{d^2\eta\epsilon_d}\right)^{1/2}
\int_{-\infty}^{\infty}dk\left(\frac{1-e^{-k^2}}{k^2}\right) \propto t^{1/2},
\end{equation}
which predicts a {\em sub-linear} time-dependence\cite{blatter1}.   
By contrast,
an uncoupled pancake 
vortex has an ordinary diffusive transverse motion, in which
$\langle |{\bf r}_{i,k}(t) - {\bf r}_{i,k}(0)|^2\rangle \propto t$.

To test this behavior, we have examined the  
long-time behavior of the quantity $\delta R(t)$
defined in eq.(\ref{eqdr}) from LD in the limits  of (i) a single pancake
vortex; (ii) a single long line; (iii) single short line segments; and (iv) an
ensemble of 64 lines in 8 and 16 layers, at various temperatures both below and
above
$T_m.$   For the single pancake, we observe ordinary diffusive behavior. From
the limit $t/t_0 \rightarrow 10^{12} $, we deduce ${\cal D} = 1.4
\times 10^{-3} {\rm cm^2/sec}$ for $T= 60 K$ (using $t_0 = 1.2 \cdot 10^{-15}
{\rm sec}$; see below).  This diffusion constant seems to agree reasonably well 
with experimental estimates for an extremely anisotropic system such as 
BiSr$_2$Ca$_2$Cu$_2$O$_8\, $\cite{song94}.  
A single line of up to 1000 vortices shows $\delta R(t)\propto t^{1/4} $
as predicted analytically, while
lines shorter than $4d$ show behaviors close to a 2D diffusion.
 
Fig.~\ref{figdr} shows case (iv) for $B = 50kG$
and 16 layers of YBa$_2$Cu$_3$O$_{7-\delta}\, $ at five temperatures
above and below $T_m$.  For $T < T_m,$ vortex excursions are limited to
radii smaller than the appropriate Lindemann distance of around $0.2a_B$.  For
$T > T_m,$ the rms displacement seems to grow with an approximately
$t^{1/4}$ behavior as expected from the analytic estimate for a single line.  
In the case
of an eight-layer system(not shown),  we observed $\delta R(t) \propto t^{1/3}$,
possibly indicating that the system  lies between the long-line and 2D limits. 
In both cases, the motion of individual pancakes is slower than in the usual
Brownian diffusion. Hence, in the long time limit, 
a diffusion constant defined on the assumption of a
linear t-dependence will vanish even in the liquid regime and even without 
pinning.
At sufficiently high temperatures, of course, line cutting will eventually
set in (though not in the present approximate model).   With line cutting, 
the system may then cross over to a 2D liquid with ordinary 
diffusive behavior. Therefore, there is an interesting possibility of two
different types of liquids characterized by different diffusive behavior, 
with a smooth crossover between them.   

As a further means of studying relaxation in the solid and liquid phases,
we have monitored the LD evolution of various thermodynamic quantities, such
as various components of internal energy and the defect density,
after the system is initialized in some arbitrary non-equilibrium state.
For each temperature, we considered five
different initial states.  Each initial state was prepared by
randomly and rigidly displacing the individual vortex lines from a perfect
triangular lattice by an amount not exceeding a fraction $\alpha_r$ of the mean
intervortex spacing. In most cases, a rapid exponential relaxation of 
energy is
observed.  By fitting the time-dependent internal energy to the form 
$a + b \exp [ -t / \tau_r ]$, we find a relaxation time $\tau_r$ which
increases with increasing temperature (cf.\ Fig.~\ref{figrelax}, for which
$\alpha_r = 0.31$). This increase is once again due mainly to an
energy scale which diminishes [or a $\lambda_{ab}(T,B)$ which increases] with 
increasing $T$.  

Superimposed on this overall trend, 
there may be a peak in $\tau_r$ corresponding to 
enhanced fluctuations precisely at the melting temperature (although there
are large uncertainties at this temperature). 
This behavior is consistent with expectations for a phase transition
with a kinetic or ``critical" slowing down, but the feature is generally 
concealed by the steep increase of $\tau_r$ with increasing temperatures,
and can hardly be used to
distinguish between possible first-order and continuous phase transitions. 
The middle row of Fig.~\ref{figrelax} shows the progression of the
density correlation function from an initial random configuration to a well
ordered state over a period of about
$100 t_0 $ at a temperature $T/T_m = 0.9$.   
The lower row shows the local instantaneous magnetic field
profile for the initial and final configurations in a particular layer. 
The cores of the vortices lie at the centers of the bright regions.

In Fig.~\ref{figrelaxnopin}, we show a typical relaxation of total internal
energy over time after an initial randomization, 
this time with $\alpha_r = 0.63$.
Also shown is the density of isolated
disclinations, $n_d^*$.  In contrast to the internal energy, which relaxes
exponentially, the disclination
density relaxes toward zero (i.\ e., an elastic lattice) with
a roughly $\ln t$ behavior. The disclinations finally disappear 
(via pair annihilation)
long after the internal energy has nearly equilibrated.
The spikes in
$n_d^*$ on top of the overall logarithmic decay may correspond to a 
large-scale rearrangements of vortices, possibly involving activated 
processes.   
The energy relaxation itself shows no obvious signature of any such activated
processes.  

To understand real materials, 
we next examine how point pins influence 
these general features.   
In Fig. \ref{figrelaxall}, we show the relaxation of
the in-plane component of the internal 
energy for different pinning strengths ($0 < \alpha_p < 5$) with an
areal density equivalent to $B_p = 888$ kG.
The relaxation is qualitatively different from
the weak-pinning cases ($\alpha_p \le 3$), in which the
energy typically decays exponentially as in clean systems. 
At $\alpha_p = 5,$ for example, the in-plane part of the energy varies
logarithmically with time almost from $t = 0$.

For pins of any strength, we find that the system 
never relaxes back to the perfect triangular lattice within our simulation
times; a significant fraction of 
disclinations always survives even after $t/t_0 = 5 \times 10^5$
(cf.\ lower part of Fig. \ref{figrelaxdefect}).
Possibly, early-stage relaxation is controlled mainly by 
pinning forces.  But once the
pinning energy is nearly optimized by this quick relaxation,
all driving forces for relaxation
become comparable in strength (after $t/t_0 \approx 3 \times 10^4$), and
relaxation slows dramatically.  Any further relaxation must reduce the
elastic strain energy without sacrificing pinning energy.  To accomplish this,
both total and isolated defect densities slowly but steadily 
decrease (with many fluctuations).  Also,  
$n_d^*/n_d$ [lower panel of Figure] decreases noticeably for
$t/t_0 > 3 \times 10^4$.   In other words, energy relaxation in this regime
is proceeding via a decrease in the number of isolated defect configurations. 

The slow elastic relaxation just described must involve vortex rearrangement. 
It may therefore affect measurements of the local magnetic field and 
the total magnetization.  Indeed, such quantities have long been known to
depend strongly on the time scale over which they are measured.
Thus, our model can help to explain such slow relaxation of magnetization, in
disordered samples.  The calculations also suggest that the slow relaxation
results predominantly from the rearrangement and annealing-out of topological
defects.  

Next, we briefly discuss the value of $t_0$, which is crucial in connecting 
our numerical results to experiments on real materials.   The
friction coefficient $\eta$ is related to the flux flow resistivity
$\rho_{ff}$ by the equation $\rho_{ff} = \frac{B \phi_0}{\eta c^2}.$
Assuming a flux-flow resistivity
$\rho_{ff} ({\rm B=50 kG,T/T_m=0.9}) \sim 0.16\, \rho_n$, where $\rho_n$ is the
normal state resistivity, and estimating $\rho_n$ as its value at $T_c(0)$, 
$\rho_n(T_c(0)) \sim 100 \mu\Omega$-cm\cite{iye88} for a single crystal 
of YBa$_2$Cu$_3$O$_{7-\delta}\, $, we obtain $\rho_{ff} \approx 1.6 \times 10^{-17}$ sec.  The
corresponding value of $t_0$, assuming the time-step in Sec.\ IIC, is
$t_0 \approx 10^{-14} - 10^{-15}$ sec.
Since our value of $\rho_{ff}$ is probably an overestimate, this 
value should be taken as the lower bound for $t_0$\cite{caveat}. 

Finally, we briefly return to the influence of pins on the dynamics
of flux lines, in light of this value for $t_0.$  In real materials, such pins
will strongly affect flux line dynamics, typically slowing down their
motion by several orders of magnitudes (as we have noted above).
This has some implications for
interpreting the experimental results which are sensitive 
to the time evolution of
local magnetic field.  One such experiment is the
spin-echo NMR probe\cite{song94,moonen94} in the highly anisotropic Tl-based
compounds.   This
has been interpreted as implying a vortex diffusion constant 
${\cal D}$ of $10^{-4}
\sim 10^{-5} {\rm cm^2/sec} $ at temperatures
$T \stackrel{<}{\scriptstyle \sim}  T_m$, and of 
$\approx 10^{-3}$ - $10^{-4} {\rm cm^2/sec} $ for $ T > T_m$.  
On the other hand, more recent spin-echo results on aligned powders of
YBa$_2$Cu$_3$O$_{7-\delta}\, $
seem to show lack of diffusion 
even on very long time scale\cite{pennington95,song95}.  These results may
indicate the presence of strong pins in the powder samples, which greatly slow
down the dynamics.  Another possibility is that the discrepancy is
somehow due to the nondiffusive behavior of long lines discussed above.

\section{Discussion and Conclusions.}

We have presented a detailed numerical study of the vortex structure in
a model for YBa$_2$Cu$_3$O$_{7-\delta}\, $, both above and below the flux lattice melting temperature,
using Monte Carlo and Langevin simulations.   Our results indicate
that in clean YBa$_2$Cu$_3$O$_{7-\delta}\, $, there is a weakly first-order melting transition,
with a very small heat of fusion which agrees the calculations of 
\cite{sasik95} at similar fields.  

We also find that there is a striking
change in distribution of local magnetic fields at melting: the time-averaged
magnetic field sensed by a fixed particle in the sample has a very narrow
distribution in the liquid (because the vortices move around in the liquid),
but acquires an asymmetric distribution of finite and temperature-dependent
width in the solid phase.  This behavior
agrees with both $\mu$SR and NMR experiments in YBa$_2$Cu$_3$O$_{7-\delta}\, $.
Although some previous MC
calculations\cite{schneider95} have suggested a similar transition, our LD simulations
provide more information.  In particular, because the time constant of the Langevin
simulations can be extracted from measurements of the liquid-state resistivity, we can
estimate the measurement time necessary to distinguish the solid and liquid field
distributions. This time appears to be about 0.5 $\mu$sec in YBa$_2$Cu$_3$O$_{7-\delta}\, $ at a field of 50 kG.

Our calculations show that there is a qualitative change in the structure
of the topological defects in the crystalline phase just below melting.
Namely, even in clean YBa$_2$Cu$_3$O$_{7-\delta}\, $, there are long lines of disclinations parallel
to the $c$ axis at relatively low fields (B $\stackrel{<}{\scriptstyle \sim}
 10 kG$), but these
break up into short (``2D'') disclinations at higher fields.  We believe that
this breakup may be an essential ingredient in the ``3D-2D'' transition 
recently reported in BSCCO\cite{cubitt93,obara95,ryu96}.
In clean systems, we find numerically that the distribution of defect line 
lengths is approximately exponential in the liquid, but deviates from 
exponential in the solid phase. We account for these forms
by a simple model of topological defects in the vortex lattice of a layered
superconductor. 

A remarkable result of our Langevin simulations is that the vortex lines
move nondiffusively even in the liquid state - that is, the rms displacements
of the vortex pancakes vary like $t^{\alpha}$, where $\alpha <  0.5$.   
This agrees with the exact result derived for a 
single long line\cite{blatter1}, 
that the rms displacement of a sufficiently long 
vortex line subject to thermal noise 
varies as $t^{1/4}$. 
Our simulations suggest that the behavior is also found in
a dense line liquid with strong short range repulsive interactions. 
While the full  transport implications remain to be explored, this 
result is  consistent with recent NMR measurements on
YBa$_2$Cu$_3$O$_{7-\delta}\, $\cite{song95} which also suggest that in the liquid state
vortex lines move subdiffusively.

The LD calculations also reveal that the vortices relax much more slowly
as $T \rightarrow T_{c2}(B)$.  Such slow relaxation is actually built into
the TDGL equation, whose time constant is proportional to the deviation of
the Ginzburg-Landau free energy from its minimum value.  Since the minimum
becomes ever shallower as T$_{c2}(B)$ is approached, the relaxation time
constant should become ever longer.  It would be most interesting to have
experimental confirmation of this behavior.  The Langevin results also
suggest that the time constant may increase as the (first-order) melting
transition is approached from either side.    

A final conclusion has to do with the influence of point pins on the dynamics. 
Our Langevin results show that, after any disturbance, the energy relaxes 
nearly exponentially back to equilibrium in a pin-free sample.   
With strong pins present, however, the energy relaxation is close to 
logarithmic in time.  At even longer time scales, there is a crossover to 
a logarithmic relaxation with a different slope.  We identify this
slower relaxation with the logarithmically slow 
annealing out of topological defects in the 
disordered samples (these same defects are also slow to
disappear in the  pin-free case).
It has long been
known that many properties of the  high-T$_c$ 
materials vary logarithmically slowly with
time  (most notably, the sample magnetization).  Our model, which treats in a
fairly realistic manner a disordered YBa$_2$Cu$_3$O$_{7-\delta}\, $ sample, clearly produces such
slow relaxation, and identifies it with a particular type of topological
defect dynamics.  
 
In summary, we have presented a detailed study of the 
nature of the flux line melting based on the vortex representation of the 
Lawrence-Doniach model, using both Monte Carlo and Langevin simulations in a
consistent way.   We calculate a wide range of properties of both the vortex
solid and liquid phases which are consistent with experiment, and which
shed new light on the topological defects which underlie the melting process,
as well as the time-dependent magnetization and magnetic field distribution,
of these materials.  

\section{Acknowledgments}

We are grateful for valuable discussions with Professor C. H. Pennington
and C. Recchia.
This work was supported by DOE Grant DE-FG02-90 ER45427
through the Midwest Superconductivity Consortium at Purdue University, by NSF
Grant DMR94-02131, and by an Ohio State University Postdoctoral Fellowship
awarded to SR.  DS thanks the Department of Applied Physics at
Stanford University, and Professor S. Doniach, for their kind hospitality
during the completion of this work.  
Calculations were carried out, in part, with the use of the
computational facilities of the Ohio Supercomputer Center.

\begin{figure}
\caption [first figure]
{
Probability distribution ${\cal P}(U)$
for the total internal energy $U_{total}(T)$ per pancake,
averaged over two different time windows (a) and (b) 
(specified in Fig.~\ref{figtwo}), at the melting temperature
$T_m$.   Calculation is carried out for YBa$_2$Cu$_3$O$_{7-\delta}\, $ at 
B = 50{\rm kG}.  $U_{hex}$ is the
corresponding energy for a perfect triangular lattice of straight vortex
lines (the minimum energy configuration).   The distributions in each
window are obtained by dividing the data from the Monte Carlo configurations
shown in Fig.~\ref{figtwo} into $256$ energy bins, each of width
$\triangle U / k_BT = 1.4\times 10^{-4}$.  The curves are least-squares fits of
these data to Gaussian  distributions.  The error bars at various values of
$U_{total}$  are the rms deviations of the 
values of ${\cal P}(U)$ inferred from
ten  bins in the vicinity of $U$.  
Inset: in-plane rms displacement $c_L$ of pancake vortices 
from their equilibrium lattice positions in units of lattice spacing, as
calculated using  both Monte Carlo and Langevin techniques. }
\label{figone}
\end{figure}

\begin{figure}
\caption [second figure]
{Evolution of the in-plane component of internal energy, $U_{inplane}(T)$, 
at the melting temperature, for B = 50{\rm kG} as calculated via MC simulation.
Insets: Density-density correlation $C({\bf r},0)$, 
averaged over
the two different time windows  (b) and (c).   
These plots show that the system is alternating between lattice and liquid
phases.}
\label{figtwo}
\end{figure}

\begin{figure}
\caption [defect figure]
{Snapshot of disclination distribution in a
single layer of a vortex liquid containing 256 lines, as determined by 
Delaunay triangulation. 
Black dots: fivefold disclinations; gray dots: sevenfold disclinations.
(a) A pair of bound disclinations, equivalent to a dislocation; 
(b) an isolated disclination; (c) a
pair of bound dislocations. The bottom row shows a typical vortex line 
liquid and its representation in terms of disclinations of either
sign. }
\label{figdefects}
\end{figure}

\begin{figure}
\caption [hex figure]
{Hexatic order parameter $|\psi_6|^2$ (left scale) and 
density of isolated disclinations $n_d^*$ (right scale)
as a function of temperature in a
64-line system of thickness 8 layers with ${\rm B}=50 {\rm kG}$.}
\label{fighex}
\end{figure}

\begin{figure}
\caption [defectscrossover figure]
{Darker symbols: probability distribution of 
disclinations of length $l_d$ at
B $= 10$, $50$, and $90$ kG for a 32-layer system, at  
temperatures such that 
$\delta R(B,T) = 0.15,$ i.\ e., slightly below melting.  Lighter symbols:
corresponding probability distributions for the same fields and system
at temperatures deep into the liquid phase.
In the right column, snapshots of
typical defect configurations in the solid phase are shown for each field.
Black and gray represent disclinations of either sign.}
\label{figdefectcrossover}
\end{figure}

\begin{figure}
\caption [pbt figure]
{Dependence of time-averaged magnetic field distribution 
${\cal P}_t(B)$ on time
window $t/t_0$ used for average, at an applied field of $50kG$.  
The time evolution is
calculated at several temperatures
from Langevin dynamics using Clem's prescription for computing
the magnetic field\cite{clem91}. 
The distribution is plotted as a function of
$B/B_{av} - 1$, where $B_{av}$ is the space-averaged 
magnetic induction.  Successive distributions in each vertical panel
are displaced horizontally by 0.02 units.}
\label{figpbt}
\end{figure}

\begin{figure}
\caption [bsq figure]
{Mean-square width $(\delta B)^2/B_{av}^2$ of the magnetic field 
distribution ${\cal P}(B/B_{av})$, plotted as a function of
$T/T_m(B)$ at an applied field B = $50kG$.  
The width is calculated from Langevin dynamics for 
time intervals of varying durations: $t/t_0 = 10^3,
10^4, 10^5$ (inset), and $5 \times 10^5$ (main Figure). 
Also shown in the inset is
the mean-square width as obtained from a MC simulation 
sampled over $4\times 10^4 $ MC steps.}
\label{figbsq}
\end{figure}

\begin{figure}
\caption [dr figure]
{(a) Rms transverse displacement $\delta R(t)$
of vortex pancakes at several temperatures both above and below the
melting temperature T$_m$, 
as calculated by Langevin dynamics at B = 50kG for $N_z =16$, and plotted in
units of 
$a_B$, the mean vortex separation.  For $T<T_m$,
the rms displacement saturates at a value approaching the Lindemann
number $c_L \approx 0.18$ at T$_m$.   In the liquid
regime, the displacement increases roughly as $t^{1/4}$, as expected for a long
line, and not as $t^{1/2}$ as expected of diffusing particles. 
(b) Root-mean-square ``wandering length''
$l_T$ at the same temperatures.  Note that $l_T$ saturates even in the
liquid state, though its value continues to increase, 
primarily because of the reduction of 
line tension with increasing temperature.}
\label{figdr}
\end{figure}

\begin{figure}
\caption [relax figure]
{Top: relaxation time $\tau_r$ describing return of 
vortex system to its equilibrium
configuration, following an initial perturbation, as plotted for several
temperatures in the flux solid and flux liquid state.  Calculation is carried
out as described in the text.  Error bars are standard deviations from five
different initial configurations.  Middle: snapshots of density-density
correlation function $C({\bf r},z=0)$ 
for three representative configurations during relaxation at $T/T_m = 0.89$.
Bottom: evolution of 
local field $B({\bf r}, t)$ in a specific $ab$ plane of the sample, for the
first and third configurations of middle panel.}
\label{figrelax}
\end{figure}

\begin{figure}
\caption [relaxnopin figure]
{Time-dependence of total internal energy U$_{total}$(T) per pancake
(left-hand scale)
and the isolated disclination density $n_d^*$ (right-hand scale), 
as calculated via  
LD simulation at $B=50 kG, T/T_m = 0.898,$ starting from a randomized initial
configuration in a pin-free sample.  
U$_{hex}$ is the internal energy of an ideal hexagonal lattice
at the same temperature.}
\label{figrelaxnopin}
\end{figure}

\begin{figure}
\caption [relaxall figure]
{Relaxation of in-plane portion of internal energy per pancake, 
U$_{inplane}$, for point pins of 
varying strength$( 0 \le \alpha_p \le 5)$. The initial
configuration is a randomized distribution of straight vortex lines
with $\alpha_r = 0.63$ (see text).
Note the gradual 
crossover from an exponential to a logarithmic time dependence as the 
pinning strength increases.  At the largest value of $\alpha_p$, the
relaxation is dominated by changes in the pinning energy.  Other conditions 
same as in the previous Figure.  $U_{hex}$ is here the in-plane part of
the internal energy of an ideal hexagonal lattice (per pancake) at the
same temperature.} 
\label{figrelaxall}
\end{figure}

\begin{figure}
\caption [relaxdefect figure]
{Time-dependence of interlayer part of internal energy U$_J(T)$, 
pinning energy U$_{pin}$(T), in-plane component of elastic energy 
U$_{inplane}$(T), and ratio of isolated disclination density $n_d^*$
to total disclination density $n_d$, for
strong pinning ($\alpha_p = 5$), at a temperature of $T/T_m = 0.898$ where
$T_m$ is the melting temperature for a pin-free sample.}
\label{figrelaxdefect}
\end{figure}


\begin{references}
\bibitem{blatter1} A more general result, for manifolds of dimension
$d=1,2,3$ moving under the influence of thermal noise, has been given by 
\cite{blatter95}.
\bibitem{caveat}  This estimate is for a clean, i.\ e., a pin-free, system.
The influence of the pins could possibly 
be approximated by a renormalized effective friction coefficient.  But
since the effect of pins varies widely from sample to sample, 
it may be difficult to extract a reliable value of $\eta$ from  
experiment.
\end{references}
\end{document}